\documentclass[twocolumn, superscriptaddress, secnumarabic, amssymb, nobibnotes, aps, prd, preprintnumbers, nofootinbib]{revtex4-1}

\RequirePackage[colorlinks=true
,urlcolor=blue
,anchorcolor=blue
,citecolor=blue
,filecolor=blue
,linkcolor=blue
,menucolor=blue
,linktocpage=true
,pdfproducer=medialab
,pdfa=true
]{hyperref}

\usepackage{amsmath,amssymb,amsthm,amsfonts}
\usepackage{graphicx,tabularx}
\usepackage{color}
\usepackage{multirow}
\usepackage{comment}
\usepackage{enumitem}
\usepackage[capitalize]{cleveref}
\usepackage{slashed}
\usepackage[normalem]{ulem}
\usepackage{scalerel}
\usepackage{wrapfig}
\usepackage{cancel}
\usepackage{xcolor}
\usepackage{tikz-feynman}

\setlist[description]{leftmargin=0.3cm}
\setlist[itemize]{leftmargin=0.5cm}

\newcommand{\be}{\begin{equation} \begin{aligned}}
\newcommand{\ee}{\end{aligned} \end{equation}}
\newcommand{\beqa}{\begin{eqnarray}}
\newcommand{\eeqa}{\end{eqnarray}}

\def\figureautorefname~#1\null{Fig.\,#1\null}
\def\tableautorefname~#1\null{Tab.\,#1\null}

\def\equationautorefname~#1\null{Eq.\,(#1)\null}

\RequirePackage[normalem]{ulem}

\crefname{section}{Sec.}{Secs.}
\crefname{figure}{Fig.}{Figs.}
\crefname{equation}{Eq.}{Eqs.}
\crefname{table}{Table}{Tables}
\crefname{appendix}{Appendix}{Appendices}

\definecolor{nicegreen}{rgb}{0., 0.75, 0.46}

\begin{document}
   

\title{{\Large Muon-induced di-tau production as a probe of new physics}}

\author{Sudipta Das}
\email{sudipta-das@uiowa.edu}
\affiliation{Department of Physics and Astronomy, University of Iowa, Iowa City, IA 52242, USA}

\author{Luke Kupari}
\email{luke-kupari@uiowa.edu}
\affiliation{Department of Physics and Astronomy, University of Iowa, Iowa City, IA 52242, USA}

\author{Mary Hall Reno}
\email{mary-hall-reno@uiowa.edu}
\affiliation{Department of Physics and Astronomy, University of Iowa, Iowa City, IA 52242, USA}

\author{Sebastian Trojanowski}
\email{sebastian.trojanowski@ncbj.gov.pl}
\affiliation{National Centre for Nuclear Research, Pasteura 7, Warsaw, PL-02-093, Poland}
\affiliation{Astrocent, Nicolaus Copernicus Astronomical Center Polish Academy of Sciences, ul.~Rektorska 4, 00-614, Warsaw, Poland}

\begin{abstract}
The muon trident process, which involves coherent scattering to produce tau pairs, is a powerful tool for constraining dark sectors. We propose to explore this channel using future high-energy muon beams with the dedicated active-target detector, High-Energy Muon Electronic Research Apparatus (HEMERA). The detector would complement muon beam-dump searches by resolving prompt signatures and could operate from preparatory facilities involving an individual muon beam of the full Muon Collider. We illustrate its capabilities for a leptophilic scalar with Yukawa-like couplings that can mediate thermal dark matter production. With $10^{18}$ incident TeV-scale muons, even a compact 10 kg silicon-based HEMERA will extend current $B$-factory and $(g-2)_\mu$ bounds by over an order of magnitude in coupling strength. In the leptophilic thermal dark matter paradigm, this search offers a way to probe dark sector targets beyond the neutrino floor of dark matter direct detection. This provides an additional compelling motivation for the development of high-energy muon beam infrastructure.
\end{abstract}

\maketitle 

\section{Introduction}
\label{intro}

The use of high-energy muon beams in future accelerator programs has gained renewed attention in recent years, particularly in connection with the proposed Muon Collider (MuC)~\cite{Delahaye:2019omf,InternationalMuonCollider:2024jyv,InternationalMuonCollider:2025sys}. The MuC promises both significantly reduced synchrotron energy losses compared to electron-positron colliders and a cleaner environment for new physics searches and precision Standard Model (SM) measurements than is possible with hadronic collisions. However, realizing this ambitious project requires technological advancements that must first be validated in dedicated experimental facilities. This preparatory phase would benefit significantly from defining independent physics goals that underscore the unique advantages of using TeV-scale muon beams.

In this context, recent discussions have centered on leveraging high-energy muon beams for various applications. These include a beam-dump search for Beyond the Standard Model (BSM) species at the MuC~\cite{Cesarotti:2022ttv,Cesarotti:2023sje,Klest:2025cnd}, new physics investigations at the proposed $\mu$TRISTAN collider~\cite{Calibbi:2024rcm}, and even the utilization of TeV-scale muons abundantly produced in the forward region of the Large Hadron Collider (LHC)~\cite{Ariga:2023fjg,MammenAbraham:2025gai,Batell:2024cdl,Francener:2025pnr,Francener:2025wzh}. This complements the existing NA64-$\mu$ experimental program, which uses a beam of $160~\textrm{GeV}$ muons~\cite{Sieber:2021fue,NA64:2024klw,NA64:2024nwj}, and various other future detectors designed to operate at similar or lower energies~\cite{Chen:2017awl,Kahn:2018cqs,GrillidiCortona:2022kbq}.

Precision muon physics at the TeV energy scale unlocks new search opportunities currently inaccessible at lower energies. A compelling example is the anticipated future measurement of the muon trident process on a nuclear target $A$ \cite{Brodsky:1966vh,Bjorken:1966kh,kelner,Henry:1968ab,Tannenbaum:1968zz,Ganapathi:1978qm,Barger:1979eg,Ganapathi:1979dj,abgs,abs,Ganapathi:1980wh,Kelner:2000va,Bulmahn:2008fa,Abbiendi:2024swt}. Here we focus on di-tau production via $\mu A\to \mu \tau^+\tau^- A$. Sample diagrams are shown in  \cref{fig:Feynman}. While the process remains unobserved, its SM cross section is predicted to grow by several orders of magnitude as the incident muon energy increases from $E_\mu \sim 100~\textrm{GeV}$ to $1~\textrm{TeV}$~\cite{kelner,abs,Ganapathi:1980wh,Bulmahn:2008fa}. Furthermore, the presence of final-state tau leptons with $E_\tau\gtrsim 100~\textrm{GeV}$ may render their short charged tracks detectable, providing an additional handle for their identification. Consequently, high-resolution experiments utilizing intense, energetic muon beams are uniquely positioned to carry out this measurement.

In this study, we consider this proposed measurement as a novel laboratory for new physics searches. Muon-induced di-tau production has recently been identified as a powerful tool for constraining dark matter (DM) particles coupled to the SM through lepton flavor violating (LFV) operators involving second and third generation leptons~\cite{Batell:2024cdl}. The sensitivity in that scenario primarily relies on the LFV-driven muon charge flip in the detector. Here, we extend this potential by demonstrating that the considered muon trident process can also probe new physics without LFV interactions, even when SM backgrounds are significant. This search could be performed using the proposed active-target experiment, hereafter referred to as the High-Energy Muon Electronic Research Apparatus (HEMERA).\footnote{In Greek mythology, Hemera was the goddess of daylight; corresponding to the experiment's ability to ``bring to light'' new physics sectors currently obscured by SM backgrounds.} We illustrate this through the study of a leptophilic scalar $S$ that exhibits Yukawa-like couplings to all lepton generations, a pattern naturally consistent with Minimal Flavor Violation (MFV)~\cite{Buras:2000dm,DAmbrosio:2002vsn,Cirigliano:2005ck}.\footnote{A Python script for SM and BSM cross section calculations and kinematic cut analysis relevant for the project is available in the taupair\_trident directory at \url{https://github.com/LuKupari/HEMERA.git}.} We find that if $S$ also mediates interactions with DM, this novel search would place limits on thermal relic targets that are complementary to DM direct detection experiments and can reach beyond the so-called neutrino floor \cite{Billard:2013qya}, where traditional underground detectors face fundamental irreducible backgrounds.

The paper is organized as follows. In \cref{sec:ditau,sec:bsmproddecay}, we discuss muon di-tau production within the SM and BSM contexts, respectively. We outline a detection strategy in \cref{sec:experiment}. \Cref{sec:bsm} focuses on the search for leptophilic scalars, detailing the proposed experimental cuts and the resulting sensitivity projections. We conclude in \cref{sec:conclusion}.

\section{Di-tau production in the SM}
\label{sec:ditau}

\begin{figure}
 \centering
    \includegraphics[width=0.49\linewidth]{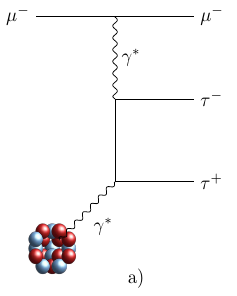}
    \includegraphics[width=0.49\linewidth]{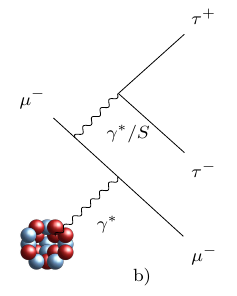}
    \caption{Sample dominant Feynman diagrams for muon-induced di-tau production in the Standard Model and new physics scalar $S$ production (on-shell) in dark bremsstrahlung followed by its decay.}
    \label{fig:Feynman}
\end{figure}

Muon di-tau trident processes in the SM can proceed through coherent scattering with the nucleus, incoherent scattering with the constituent nucleons, inelastic scattering with the nucleons and scattering from atomic electrons. Unlike in $e^+ e^-$ pair production in $\mu e\to \mu e^+ e^- e$, muon production of $\tau^+\tau^-$ is very small relative to the other processes in the energy range relevant to this paper \cite{Bulmahn:2008fa}. As further discussed in \cref{sec:experiment}, we focus on detection of coherent scattering with the nucleus.

There is an extensive literature on muon initiated  trident      processes \cite{Brodsky:1966vh,Henry:1968ab,kelner,abgs,abs,Ganapathi:1980wh,Ganapathi:1979dj,Barger:1979eg,Kelner:2000va,Bulmahn:2008fa}. \Cref{fig:Feynman} shows representative Feynman diagrams for di-tau production for coherent muon scattering with $A$. In the SM, di-tau production in coherent scattering with $A$ can proceed through the sub-process of virtual light-by-light scattering, and by a virtual photon conversion to di-tau pairs from either the initial- or final-state muon. $Z$-boson mediated diagrams are heavily suppressed in the kinematic regime of interest. For coherent scattering, the momentum transfer to the nucleus is small, $|q^2|\ll M_Z^2$, and the typical invariant mass of the $\tau^+\tau^-$ pair remains well below the $Z$ pole, rendering the weak contribution negligible. Finally, the process where a virtual photon is emitted directly from the nucleus to convert into a di-tau pair is also negligible compared to the first two mechanisms and will not be discussed further.

The first two processes are represented by the ``a'' (Bethe-Heitler-like) and ``b'' (Compton-like) diagrams in \cref{fig:Feynman}. The two sets of diagrams are separately gauge invariant. The cross section from the a-diagrams dominates the total cross section because there are two space-like photon propagators, whereas in the b-diagrams, there is one space-like and one time-like photon propagator. The interference term between a- and b-diagrams vanishes in the total cross section because of charge-conjugation invariance
\cite{Brodsky:1966vh,Henry:1968ab,Ganapathi:1980wh}.
The a-diagrams with $\gamma^*\gamma^*\to \tau^+\tau^-$ are C-even, while the b-diagrams with $\gamma^*\to \tau^+\tau^-$ are C-odd, so the interference term in the total cross section vanishes. Indeed, for differential distributions in which  the taus of both charges are treated uniformly, the interference of a- and b-diagrams does not contribute \cite{Brodsky:1966vh,Henry:1968ab}. 
We include both a- and b- contributions in what follows, and we consider distributions for which the interference diagrams do not contribute.

For the process 
\begin{equation}
\mu(k)+A(p)\to \mu(k_1)+\tau^+(p_+)+\tau^-(p_-)+A(p_2)
\end{equation}
we define $q= p-p_2$, $q^2=-Q^2$, and $p^2=p_2^2=M^2$. 
For a-diagrams, the photon propagator $q_a=k-k_1$ connects the muon propagator to the tau pair, while for the b-diagrams, this photon propagator has momentum $q_b=p_++p_-$. We define $q_b^2= m_{\tau\tau}^2$. The differential cross section can be written as
\begin{eqnarray}
\nonumber 
    {\rm d}\sigma &=& \frac{(2\pi)^4}{2\sqrt{(2p\cdot k)^2-4m_\mu^2 M^2}}\sum_{d=a,b}L_{\mu\nu}^{(d)}W^{\mu\nu} \frac{e^8}{q^4\, q_d^4}\\
    \nonumber
&\times &    \delta^4(k+p-k_1-p_+-p_--p_2)\\
    &\times & \frac{{\rm d}^3 k_1}{2E_1 (2\pi)^3}\,
    \frac{{\rm d}^3 p_+}{2E_+ (2\pi)^3} \, 
    \frac{{\rm d}^3 p_-}{2E_- (2\pi)^3}\frac{{\rm d}^3 p_2}{2E_2 (2\pi)^3}
\end{eqnarray}
The leptonic tensors for $d=a,b$ can be written respectively in terms of tensors for the muon traces and the tau traces \cite{kelner,abs} (see also refs. \cite{Bulmahn:2008fa,Bulmahn:2010qna}):
\begin{eqnarray}
L_{\mu\nu}^{(a)}&=& A^{(\mu)}_{\alpha \beta}[B^{(\tau)}]^{\alpha \beta}_{\mu\nu}\\
    L_{\mu\nu}^{(b)}&=& C^{(\tau)}_{\alpha \beta}[D^{(\mu)}]^{\alpha \beta}_{\mu\nu}\,.
\end{eqnarray}
The hadronic tensor can be reduced to
\begin{equation}
\label{eq:wmunu}
       W^{\mu\nu}=  -g^{\mu \nu} W_1(Q^2)
    + \frac{p^\mu p^\nu}{M^2} W_2(Q^2)\,.
 \end{equation}
We used the symbolic manipulation program FORM to evaluate the matrix elements \cite{Ruijl:2017dtg}.

The nuclear form factors, with the definitions and normalizations above, can be written as 
\begin{eqnarray}
 \frac{W_1}{4M^2} &=& \tau G_M^2 \\
 \frac{W_2}{4M^2} &=& \frac{G_E^2+\tau G_M^2}{1+\tau}\\
 \tau &=& \frac{Q^2}{4M^2}\,.
 \label{eq:form_factors}
\end{eqnarray}
For a proton target, $G_E(Q^2)=G(Q^2)$ and $G_M(Q^2) = 2.79\, G(Q^2)$, where we use the dipole approximation for $G(Q^2)$ \cite{Perdrisat:2006hj}:
\begin{equation}
    G(Q^2) = \frac{1}{(1+Q^2/0.71\ {\rm GeV}^2)^2}\,.
\end{equation}
More sophisticated treatments of the free nucleon form factor with the z-expansion method are described in refs. \cite{Lee:2015jqa,Ye:2017gyb}. The dipole form is sufficient for the analysis presented here.

For nuclear targets considered here, we take the magnetic form factor equal to zero, and replace the proton form factor $G_E$ with the form factor that includes electron screening, which for large $A$ can be written approximately as \cite{Andreev:1997pf}: 
\begin{eqnarray}
     G_E^{Z,{\rm eff}}(Q^2) &=& G_E^Z(Q^2)-F_e^Z(Q^2)\\
     G_E^Z(Q^2) &=& \frac{Z}{(1+a^2 Q^2/12)^2}\\
     \nonumber
&& a = (0.58+0.82 A^{1/3})5.07\ {\rm GeV}^{-1} 
\\
F_e^Z (Q^2) & = & \frac{Z}{1+a_e^2Q^2}\\
\nonumber
&&a_e = 111.7 Z^{-1/3}/m_e\, .
\label{eq:formfactornuclear}
\end{eqnarray}
We have assumed the nuclear target is a fermion. The results are similar with a scalar nuclear target \cite{Lindner:2016wff,Chu:2018qrm}. We use the adaptive Monte Carlo code VEGAS to perform the numerical integrations \cite{PETERLEPAGE1978192}.

\section{Leptophilic scalar production and decay}
\label{sec:bsmproddecay}

The muon di-tau trident signature can also be mediated by BSM interactions, opening a new window to constrain new physics. For instance, this process can be driven by heavy new physics parameterized through dimension-six SMEFT four-lepton operators involving the $\mu$--$\tau$ sector that are notoriously difficult in $e^+e^-$ colliders~\cite{Singh:2026xak}. Below, we illustrate this by considering the search for leptophilic scalars $S$ with Yukawa-like couplings. The relevant interaction term with the SM leptons $\ell$ is described by
\begin{equation}
\mathcal{L} \supset -\xi\sum_{\ell=e,\mu,\tau}{\frac{m_\ell}{v}\,S\ell\bar{\ell}},
\label{eq:Lagrangian}
\end{equation}
where $m_\ell$ are the lepton masses, $v=246~\textrm{GeV}$ is the SM Higgs boson vacuum expectation value (vev), and the dimensionless coupling strength is denoted by $\xi$. The UV completion of the operator in \cref{eq:Lagrangian}, which preserves SM gauge invariance, might arise, e.g., from a lepton-specific two Higgs doublet model~\cite{Batell:2016ove}.

The Yukawa-like structure of couplings is inspired by the MFV framework, which relates flavor-changing BSM patterns to the relevant SM structure~\cite{Buras:2000dm,DAmbrosio:2002vsn,Cirigliano:2005ck}. We note, however, that even stronger, or modified, hierarchy between the couplings of the new scalar can be obtained if this assumption is lifted~\cite{Batell:2017kty}. The considered muon-induced trident process would remain a useful probe of such scenarios as long as both couplings to muons and tau leptons are not too strongly suppressed.

Light new scalars with masses below the SM Higgs boson mass can induce a substantial BSM contribution to the anomalous magnetic moment of the muon, $(g-2)_\mu$~\cite{Kinoshita:1990aj}, cf. also Refs~\cite{Chen:2015vqy,Batell:2016ove} for discussion specific to leptophilic scalars. This prompted further analyses proposing the search for such new scalars in active beam-dump experiments~\cite{Chen:2017awl,Chen:2018vkr}. The relevant bounds apply primarily to invisible decay modes of these scalars, or to scenarios with displaced $S$ decays, cf. Ref.~\cite{NA64:2024nwj} for the recent NA64-$\mu$ constraints.

The invisible decay mode may arise naturally in scenarios where $S$ mediates interactions between the SM and dark matter (DM), e.g.,
\begin{equation}
\mathcal{L}\supset -g_\chi S\chi\bar{\chi},
\end{equation}
where we have introduced a new Dirac fermion $\chi$ as a DM candidate. However, for $m_\chi > m_S$, the scalar decays visibly into SM leptons even in this scenario. In this visible decay case, and for $g_\chi \gg \xi m_\tau/v$, the DM relic density is set by secluded annihilations, $\chi\bar{\chi}\to SS$, independently of the $\xi$ coupling strength. The relevant dark coupling constant can be found by fitting the observed DM abundance~\cite{Planck:2018vyg}, and it corresponds to $g_{\chi,\textrm{th.}} \sim 0.1\times\sqrt{m_\chi/10~\textrm{GeV}}$ for $m_\chi \gg m_S$~\cite{Batell:2018fqo}. The secluded annihilation mode is $p$-wave suppressed, thereby evading stringent bounds from DM indirect detection. 

Regarding DM direct detection, tree-level couplings to quarks are absent and interactions with electrons are strongly suppressed. However, DM-nucleon scattering is induced at the loop level via two-photon exchange mediated by lepton triangle loops~\cite{Kopp:2009et,Garani:2021ysl}. In this case, besides electron loops, Yukawa-like couplings of the leptophilic scalar under study yield additional non-negligible contributions from heavier leptons. This elevates the effective DM-nucleus cross section to values that can be probed by modern underground detectors~\cite{Graziani:2026noc}. Searching for the scalar mediator at accelerator facilities thus offers a complementary probe to DM direct detection. By directly targeting the unstable mediator particle, such searches provide a window to test this DM portal both within and beyond the thermal DM paradigm.

In the following, we will focus on the search for a new scalar with the mass above the di-tau threshold, $m_S>2\,m_\tau$. Such a massive new scalar decays predominantly into tau pairs with a nearly $100\%$ branching fraction, while its decay widths into lighter leptons remain suppressed by factors of $m_{e(\mu)}^2/m_{\tau}^2$. The $S\to\tau\bar{\tau}$ decay length is tiny,
\begin{eqnarray}
\nonumber 
\bar{d}_{S\to\tau\bar{\tau}} &\simeq & (6.4\times 10^{-8}~\textrm{cm})\times  \left( \frac{1}{\beta_\tau^\ast}\right)^{3}\\
 & &\times   \left(\frac{4~\textrm{GeV}}{m_S}\right)^2\left(\frac{1}{\xi}\right)^2\left(\frac{E_S}{100~\textrm{GeV}}\right),
\end{eqnarray}
where $\beta_\tau^\ast = \sqrt{1-4\,m_\tau^2/m_S^2}$ is the tau lepton velocity in the $S$ rest frame. The BSM scalar decays immediately after being produced in the active target, leaving no detectable vertex separation in the considered detector setups.

\begin{figure}
 \centering
    \includegraphics[width=\linewidth]{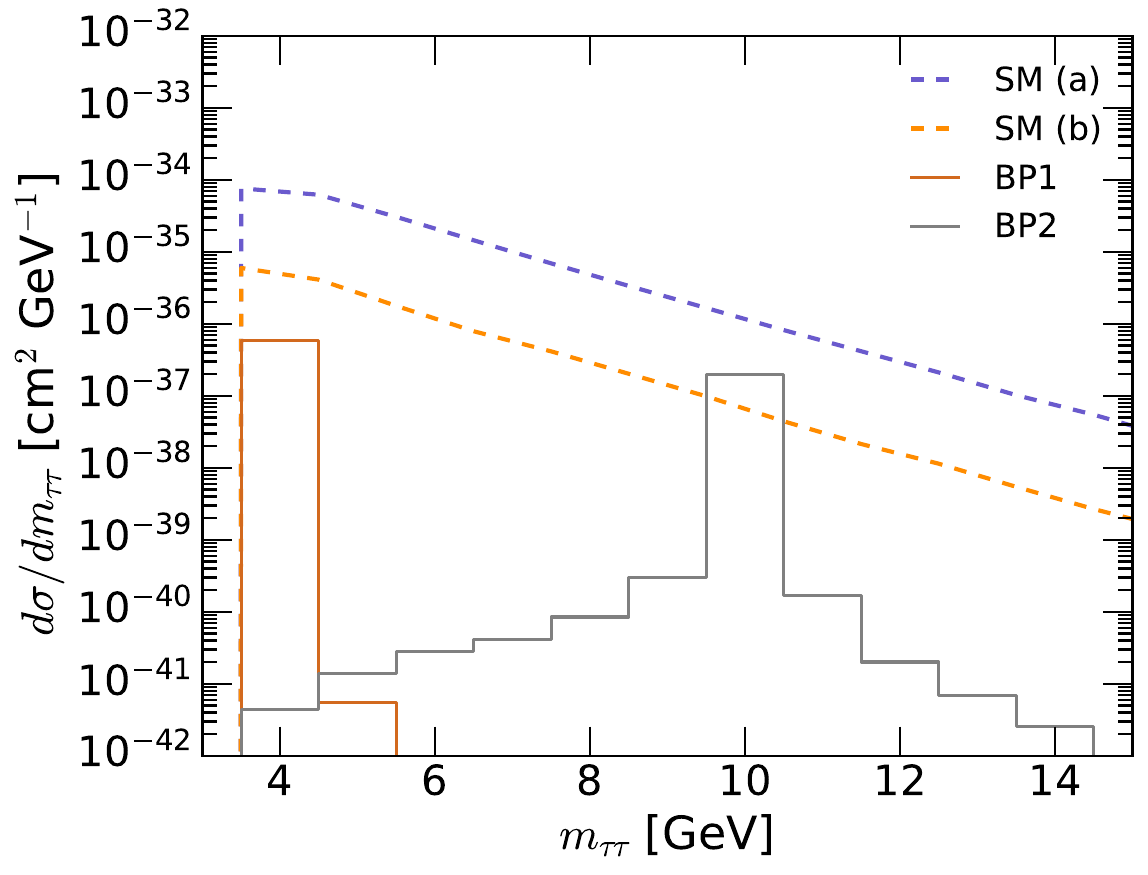}
    \caption{Differential cross section as a function of invariant mass of $\tau^+\tau^-$ pair for $E_\mu=1$ TeV. The dashed lines show the distributions for the SM processes, while the solid lines correspond to the BSM case. For the BSM process, the
    benchmark point BP1 has $m_S=4$ GeV and $\xi=5$. BP2 has $m_S=10$ GeV and $\xi=20$.}
    \label{fig:inv-mass-dist}
\end{figure}

As illustrated in \cref{fig:Feynman}, the BSM scalar $S$ is produced on-shell via high-energy muon bremsstrahlung, $\mu N \to \mu NS$, followed by the decay $S\to\tau^+\tau^-$. The differential cross section of the b-diagram with scalar $S$ is written as
\begin{eqnarray}
{\rm d}\sigma &=& \frac{(2\pi)^4}{2\sqrt{(2p\cdot k)^2-4m_\mu^2 M^2}}L_{\mu\nu}^{S}W^{\mu\nu}\\
\nonumber 
   &\times&  \frac{e^4\, g^2_\mu g^2_\tau}{q^4\,\left[(m^2_{\tau\tau}-m^2_S)^2+m^2_S\Gamma^2\right]}\\
   \nonumber
&\times &    \delta^4(k+p-k_1-p_+-p_--p_2)\\
\nonumber
    &\times & \frac{{\rm d}^3 k_1}{2E_1 (2\pi)^3}\,
    \frac{{\rm d}^3 p_+}{2E_+ (2\pi)^3} \, 
    \frac{{\rm d}^3 p_-}{2E_- (2\pi)^3}\frac{{\rm d}^3 p_2}{2E_2 (2\pi)^3}\,,
\end{eqnarray}
where $g_\alpha = \xi\,m_\alpha/v$~$(\alpha = \mu,\tau)$, and $m_{\tau\tau}$ is the invariant mass of the $\tau^+\,\tau^-$ pair. The leptonic tensor $L^S_{\mu\nu}$ includes the production and decay of the scalar $S$,
\begin{eqnarray}
    L^S_{\mu\nu} = 2(m^2_{\tau\tau}-4 m^2_\tau)\times M^{(\mu)}_{\mu\nu}\,.
\end{eqnarray}
Here, we assume the coherent scattering regime, characterized by small momentum transfer to the nucleus, and consider the prompt decay of $S$ into a tau pair. The resulting signature consists of a final-state muon and two tau leptons, accompanied by negligible nuclear recoil. The nuclear form factors are the same as SM diagrams defined in \cref{eq:form_factors}. The histograms in \cref{fig:inv-mass-dist} show the invariant mass distributions of the tau pair for the BSM process involving scalar decay for two benchmark scenarios: BP1 with $m_S = 4$ GeV and $\xi = 5$, and BP2 with $m_S = 10$ GeV and $\xi = 20$. For comparison, we also show the invariant mass distributions for the SM processes corresponding to diagrams (a) and (b) in \cref{fig:Feynman}. As illustrated, the search for a leptophilic scalar signal necessitates the efficient rejection of SM backgrounds, which is discussed below. All relevant expressions, including the leptonic tensor $M^{(\mu)}_{\mu\nu}$ not explicitly shown here, can be found in the Python script associated with this paper (see footnote~2).

\section{Detecting muon-induced di-taus}
\label{sec:experiment}

\textbf{Backgrounds} A successful search for muon tridents requires not only an intense, high-energy $\mu$ beam but also detection strategies capable of simultaneously measuring muon energy loss and identifying both tau leptons in the final state. An additional challenge is suppressing backgrounds from diffractive and deep-inelastic scattering (DIS), which involve larger momentum transfer to the nucleus. We note that muon interactions mediated by $W$ bosons (charged-current DIS) are naturally suppressed by the large $W$ mass compared to photon mediated interactions. Nevertheless, the former processes, where the primary muon is converted into a neutrino, could mimic a significant muon energy loss if a final-state pion is misidentified as the outgoing muon. Notably, however, DIS processes typically involve multiple charged tracks from the hadronic shower. Suppression of these backgrounds should be achieved by reconstructing the primary vertex to analyze the charged-track multiplicity and other kinematic variables, cf. the discussion of a proposed neutrino trident process measurement at the LHC~\cite{Altmannshofer:2024hqd}.

Crucially, while prompt hadronic backgrounds (such as final-state pions) originate at the primary vertex, the short-lived but boosted tau leptons can be identified by their distinct displaced decay vertices. This spatial separation, which demands an experimental setup with micrometer-scale vertex resolution, is the primary handle for distinguishing the di-tau signal from prompt SM backgrounds that might otherwise mimic the kinematic signatures of the trident interaction. The effectiveness of this suppression depends significantly on the tau decay topology. While 1-prong hadronic decays may be mimicked by isolated prompt pions, the identification of 3-prong decays (e.g., $\tau^\pm\to \pi^\pm \pi^\pm \pi^\mp \nu_\tau$) provides a particularly powerful rejection tool. Reconstructing three charged tracks that converge at a secondary vertex, separated from the primary vertex, provides a robust handle for rejecting prompt hadronic backgrounds.

Another potential source of background for the displaced-vertex signature is the production of heavy-flavor mesons, specifically open charm and bottom, $\mu A \to \mu c\bar{c} X$ and $\mu A \to \mu b\bar{b} X$,
with $b\bar{b}$ suppressed relative to $c\bar{c}$. Mesons such as $D^\pm$ and  $D_s^\pm$,
have decay lengths of $c\tau \sim 100-300\,
\mu\text{m}$, which are comparable to the tau lepton~\cite{ParticleDataGroup:2026mpi}. 
However, these backgrounds can be efficiently mitigated by applying a cut on the track multiplicity at the primary vertex and studying the secondary-vertex decay kinematics. First, the dominant production mechanism for open heavy flavor in muon scattering is deep-inelastic photon-gluon fusion, which produces prompt hadronic activity at the primary vertex~\cite{Leveille:1978px,Vissani:2000rr}. Instead, the coherent di-tau trident signal leaves the nucleus intact. Thus, a track-multiplicity cut at the primary vertex within the active target may effectively suppress this deep-inelastic background.

Second, for
coherent or diffractive heavy-flavor production  \cite{Frixione:1994dv,Frixione:1995qc},
the decay kinematics provide additional discrimination. For instance, the hadronic decay $D^\pm \to K^\mp \pi^\pm \pi^\pm$, which can mimic a 3-prong tau decay, lacks a final-state neutrino. In the case of the tau lepton, the corresponding missing (neutrino) transverse momentum leads to a characteristic angular deflection (kink) relative to the reconstructed intermediate tau track, which provides an additional handle on the tau lepton identification~\cite{OPERA:2015wbl}. Semileptonic heavy-flavor decays will have missing transvere momentum; however, they can also be rejected by vetoing additional prompt electrons and muons or by identifying the final-state kaons. We also note that the notorious diffractive $D_s$ meson production, followed by the decay $D_s \to \tau\nu_\tau$, which was observed as a background in neutrino trident searches at NuTeV~\cite{NuTeV:1999wlw}, arises via charged-current $W$-boson exchange, which produces a single $D_s$ meson. Instead, muon scattering mimicking the di-tau trident production proceeds via neutral currents and leads to the production of a $c\bar{c}$ pair. The probability for both charm quarks to hadronize into $D_s$ mesons and subsequently decay into tau leptons is highly suppressed. Finally, bottom meson production is kinematically heavily suppressed by the $b$-quark mass at the considered fixed-target collision energies, and the large visible invariant mass of its decay products allows for distinguishing it from the tau signature~\cite{ParticleDataGroup:2026mpi}.

Furthermore, a fixed-target muon program faces neutrino-induced backgrounds due to muon decays in flight, though these can be vetoed in the active target by tagging the incident muons. Similarly, trident events induced by high-energy electrons or positrons, which are also present in an intense muon beam, must be rejected based on the detection of the final-state $e^\pm$. The detailed modeling and mitigation of these backgrounds will depend on the final detector configuration and should be the subject of future studies using full detector simulations. In the following analysis, we highlight these challenges to define the required detector capabilities, while focusing our quantitative study on the primary irreducible SM backgrounds originating from coherent muon-nucleus interactions with tau pairs in the final state. These processes represent the fundamental physical floor for the di-tau trident search, against which the sensitivity to new physics must be evaluated.

\begin{figure*}
 \centering
    \includegraphics[width=\linewidth]{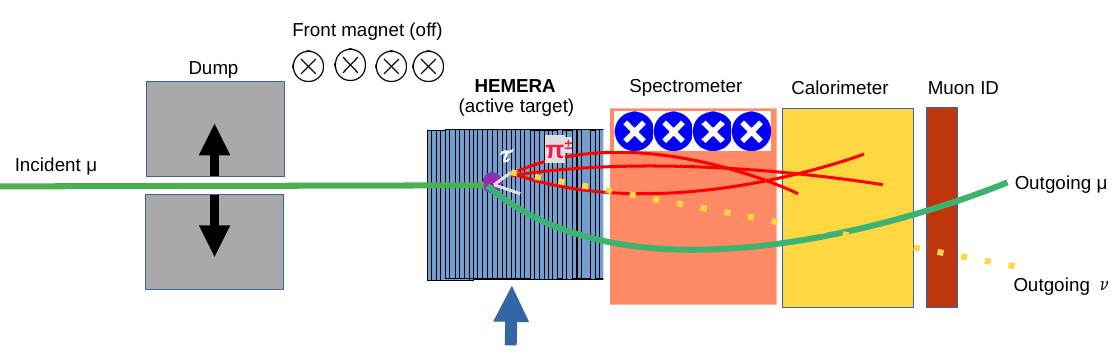}
    \caption{Schematic design of the proposed HEMERA detector. An incident muon impinges on the active target, producing two tau leptons and one outgoing muon in the final state. The tau leptons decay rapidly, leaving only short tracks in the active target material. The 3-prong tau decay into three charged pions is shown for illustration. These pions are subsequently detected in the spectrometer and deposit their energy in the calorimeter. For clarity, only the decay products of one tau lepton are depicted, although both leptons decay. The outgoing muon's momentum is measured using the spectrometer, and the muon system at the end is used for identification. The active-target operation mode of HEMERA requires lifting the front dump and moving the active target into the beam axis, as indicated by the arrows. The front magnet is off in this configuration, while its function is to deflect muons traversing the dump during the beam-dump operation mode (cf. Refs~\cite{Cesarotti:2022ttv,Cesarotti:2023sje}). The downstream detector components are based on the design of the forward neutrino detector proposed for the Muon Collider~\cite{King:1997dx}.}
    \label{fig:detector}
\end{figure*}

The main irreducible background source in new physics searches is the SM trident process, which we focus on in \cref{sec:bsm}. The signature could, however, also be mimicked by a two-step process involving a hard bremsstrahlung photon that subsequently converts into a tau pair, $\mu N \to \mu \gamma N$ followed by $\gamma N\to \tau^+\tau^- N$. This potential background is, however, heavily suppressed by three primary factors. First, in line with the experimental cuts discussed below, we require the bremsstrahlung photon to carry nearly the entire incident energy,  defined by the fractional energy transfer $\nu = E_\gamma/E_\mu\gtrsim 0.95$, which is statistically rare. The differential cross section for muon bremsstrahlung scales as $1/\nu$, leading to an interaction length for such ``catastrophic'' energy loss of approximately $70-80~\textrm{km}$ in silicon at $E_\mu=1~\textrm{TeV}$~\cite{Groom:2001kq}. For a 1-meter-long detector, the probability of such an emission is only $\sim 10^{-5}$. 

Second, the QED pair production cross section scales with the lepton mass as $1/m_\ell^2$, resulting in a suppression of more than seven orders of magnitude for tau pairs relative to the dominant $e^+e^-$ channel, given $(m_e/m_\tau)^2\approx 8\times 10^{-8}$. Finally, such a process is typically spatially displaced; a photon travels a characteristic distance governed by the radiation length ($X_0\simeq 9.37~\textrm{cm}$ in silicon ) before converting. For a vertex resolution of $\delta_z \sim 100~\mu\textrm{m}$, the probability of the photon converting close enough to the primary vertex to mimic a single coherent interaction introduces an additional suppression of approximately three orders of magnitude. These cumulative factors result in a total suppression of $\mathcal{O}(10^{-15})$ per incident muon. Although this yields a non-vanishing number of events for large muon fluxes, after kinematic cuts this two-step process remains strongly subdominant to the primary SM muon trident signal discussed below. Consequently, in the next section, we focus on the standard SM muon trident production as the primary irreducible background for this search.

\textbf{Signal detection} Having established that the muon trident process remains the main irreducible background, we now turn to the instrumentation required to isolate this signal. Reference~\cite{Batell:2024cdl} proposed expanding the concept of a muon beam-dump experiment at the MuC~\cite{Cesarotti:2022ttv} into a dedicated active-target experiment to enable the measurements above. The study showed that even a $2$ cm-thick lead plate interleaved with electronic detectors and running for a fraction of the collider time could be sufficient to probe interesting and cosmologically viable BSM scenarios. As in the beam-dump setup, the active target must be followed by a full detector to isolate the muon trident process properly. This includes a spectrometer, calorimeter, and muon ID system.

Remarkably, a similar detector setup has been proposed for the forward neutrino physics program at the MuC~\cite{King:1997dx,InternationalMuonCollider:2024jyv}; see also recent studies on BSM prospects~\cite{Adhikary:2024tvl,Kling:2025zsb}. In that context, the active target is a cylindrical vertex detector (1 m long, 10 cm radius), composed of 750 silicon tracking planes equivalent to 10 kg target material. With an average spacing between planes of approximately $1.3~\textrm{mm}$, this high-granularity detector provides a longitudinal vertex resolution $\delta_z$ of order tens of $\mu\textrm{m}$, a level of precision demonstrated by the benchmarks for modern high-granularity trackers~\cite{Barchetta:2021ibt}. This precision is more than sufficient to resolve the displaced decay vertices of the $\tau$ leptons, which at a few hundred GeV energy are boosted to characteristic decay lengths of several centimeters. Finally, this tracking density also helps identify and reject background from hard bremsstrahlung photons, as discussed above.

We schematically illustrate this experimental setup in \cref{fig:detector}, assuming it could be adapted to study high-energy muon interactions at future accelerator facilities. We refer to such a proposed detector as HEMERA. The detector would be installed in a muon beam-dump area and could operate sequentially with the proposed beam-dump experiment, as they share essential components. This sequential operation, i.e., switching between fixed-target and beam-dump modes, is a well-tested approach, used, for example, by the NA62 experiment at CERN~\cite{NA62:2023qyn}. In that case, the transition involves periodically removing the target and closing the collimator. The latter then serves as the beam dump. For the setup discussed here, the transition to the beam-dump mode would also require activating a strong magnetic field to deflect final-state muons traversing the dump, preventing contamination of the new physics search. Conversely, in the active-target (HEMERA) mode of our interest in this study, outgoing muons are an essential part of the signature and must be measured by the downstream detector components.

In this study, we focus on HEMERA utilizing a fixed-energy muon beam of $E_\mu = 1~\textrm{TeV}$. A single high-energy muon beam is sufficient to perform this search. Prior to full MuC operations, such a beam could be delivered by facilities like the proposed $\mu$TRISTAN~\cite{Hamada:2022mua}. In the following, we adopt $N_\mu = 10^{18}$ as the reference value for the number of muons on target (MOT). This corresponds to one hundred hours of operation under the baseline Muon Accelerator Program (MAP) design~\cite{MAP2,Neuffer:2018yof}.

Since the MuC is designed for even higher energies, up to a few TeV~\cite{Delahaye:2019omf,InternationalMuonCollider:2024jyv,InternationalMuonCollider:2025sys}, we also discuss results obtained for HEMERA operating at $E_\mu = 5~\textrm{TeV}$ for comparison. We adopt $10^{20}$ MOT as the reference in this case, which represents an entire MAP physics run and, therefore, corresponds to the ultimate sensitivity of the considered search~\cite{Cesarotti:2022ttv}.

With these benchmark exposures, we expect  $\mathcal{O}(10^{10})$ SM coherent di-tau trident events before kinematic cuts. As this process has never been observed experimentally, measuring it would be a valuable test of the SM in its own right. Given the expected event rates, the statistical uncertainty on the SM cross section will be strongly suppressed ($\ll 0.1\%$). We note that the incident muon beam is expected to be essentially monochromatic, with a beam energy spread of $\mathcal{O}(0.1\%)$~\cite{InternationalMuonCollider:2025sys}. Hence, the di-tau production cross section will be measured at a fixed energy, and its precision will depend on uncertainties in the total incident muon flux and tracking efficiency, combined with theoretical uncertainties on the nuclear form factors. 

Nevertheless, a measurement of the muon-induced di-tau production cross section with even a few-percent precision would be impactful. In this process, the longitudinal momentum transfer to the target nucleus scales inversely with the muon energy loss. Even for the catastrophic muon energy loss, the momentum transfer remains typically within a few MeV range. This value is several orders of magnitude greater than the typical atomic screening momentum of $\lesssim 10~\textrm{keV}$, which is relevant for dominant $e^+e^-$ production. The considered process would thus provide a clean, purely leptonic probe of the unscreened coherent nuclear form factor in the tens of MeV regime. Beyond testing the SM in this limit, the SM measurement will also facilitate calibration of the HEMERA detector for the BSM search discussed below.

Last but not least, we note that TeV-scale muons are naturally produced in the forward region of the LHC. In this scenario, an intense $\mu$ beam is only partially deflected by the strong LHC magnets and by interactions with the collider infrastructure and its environment. Consequently, a substantial number of these muons is expected to reach far-forward detectors, such as the currently operating FASER~\cite{FASER:2018ceo,FASER:2018bac} and SND@LHC~\cite{SNDLHC:2022ihg} experiments. While these muons constitute a background for typical new physics searches and neutrino measurements, they also motivate additional physics studies~\cite{Ariga:2023fjg,MammenAbraham:2025gai,Batell:2024cdl,Francener:2025pnr,Francener:2025wzh}.

We expect approximately $10^9$ MOT to hit the existing forward detectors during the ongoing Run 3 data-taking period. This number is predicted to grow by two to three orders of magnitude for the proposed Forward Physics Facility (FPF) at CERN~\cite{Anchordoqui:2021ghd,Feng:2022inv,Adhikary:2024nlv}. While limited statistics constrain the sensitivity of muon trident-based new physics searches at the LHC, existing forward neutrino detectors, such as FASER$\nu$~\cite{FASER:2019dxq,FASER:2020gpr}, already offer the capability to measure short $\tau$ lepton tracks produced in tau neutrino interactions within their tungsten targets. This provides a valuable proof-of-principle experimental setup for the proposed search. However, future, much more intense muon beams will likely require a different type of active target in HEMERA (e.g., the silicon tracking planes discussed above) to reduce detector occupancy and enable real-time reconstruction.

\section{Search for leptophilic scalars}
\label{sec:bsm}

As discussed above, the search for muon-induced di-tau tridents should rely on the joint identification of the outgoing muon and the tau-lepton pair. Given the very short decay lengths of the BSM scalar $S$, its coherent production yields a signature effectively identical to the irreducible SM trident process. Consequently, isolating the BSM signal in HEMERA depends on suppressing SM backgrounds through optimized kinematic cuts. In the following, we discuss the impact of two such cuts based on the muon energy loss and the invariant mass of the tau pair.

\begin{figure}
 \centering
    \includegraphics[width=\linewidth]{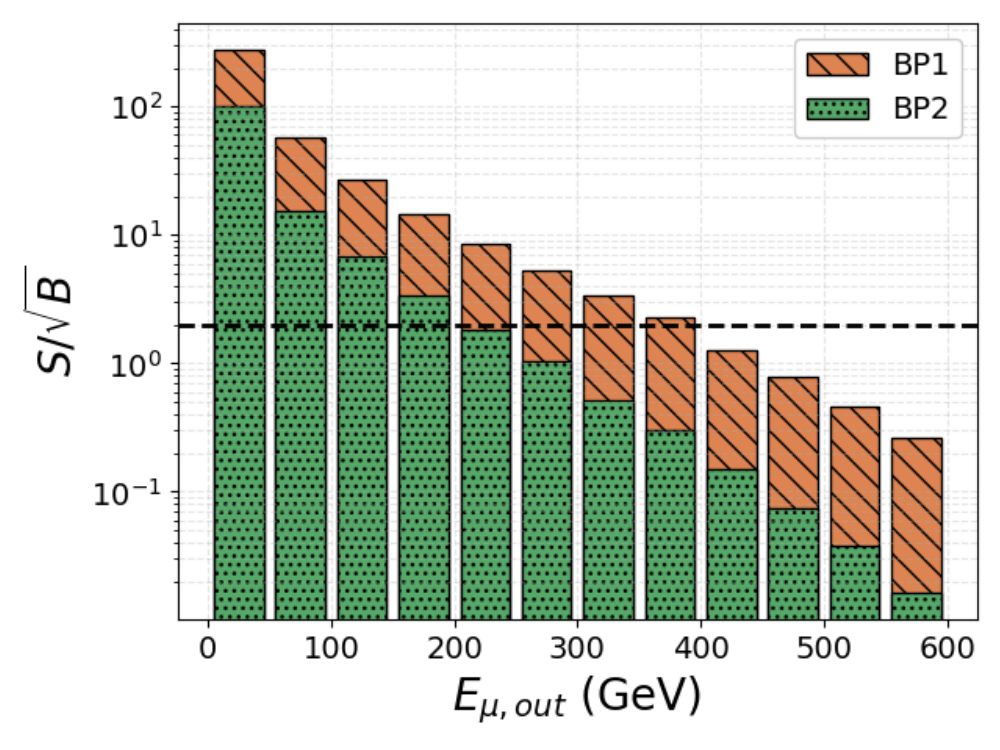}
    \caption{Signal significance $S/\sqrt{B}$ as a function of the outgoing muon energy, $E_{\mu,\textrm{out}}$, for the muon trident process in HEMERA assuming $10^{18}$ muons on target (MOT) consisting of 10 kg of silicon for incident muon energy of 1 TeV. Results are shown for the two benchmark scenarios, BP1 and BP2, where BP1 has $m_S=4$ GeV and $\xi=5$, while BP2 has $m_S=10$ GeV and $\xi=20$. The horizontal dashed black line indicates the $S/\sqrt{B}= 2$ sensitivity threshold.}
    \label{fig:muonenergyloss}
\end{figure}

\textbf{Muon energy loss} For $m_S\gg m_\mu$, the outgoing scalar tends to carry away the bulk of the incident muon energy, resulting in ``catastrophic'' energy loss for the muon. In \cref{fig:muonenergyloss}, we show the projected signal significance ($S/\sqrt{B}$) for HEMERA as a function of the outgoing muon energy, $E_{\mu,\textrm{out}}$, for a fixed incident energy of $E_{\mu,\textrm{in}} = 1~\textrm{TeV}$ and $10^{18}$ MOT. We consider two benchmark points BP1 and BP2 introduced in \cref{sec:bsmproddecay}. As illustrated, the $S/\sqrt{B}$ ratio peaks at low outgoing muon energies, as the BSM signal favors a large momentum transfer to the final-state scalar.

The efficiency of this cut depends on both the muon spectrometer's energy resolution and the ability to detect soft muons emerging from the target. Downstream magnets can significantly deflect low-energy muons before they reach the muon identification system. This presents a particular challenge for detectors optimized for high-energy interactions, which typically utilize strong magnetic fields to accurately measure ``hard'' final-state particles.

For instance, a $30~\textrm{GeV}$ threshold approximately corresponds to the current capabilities for forward muon measurements at the LHC in the FASER experiment~\cite{Feng:2017uoz,FASER:2020gpr}. Lowering this threshold, however, could significantly improve signal acceptance; for $m_S = 4~\textrm{GeV}$, the acceptance increases from $62\%$ at a $30~\textrm{GeV}$ threshold to $79\%$ at $10~\textrm{GeV}$, and $98\%$ at $1~\textrm{GeV}$. In the following, we assume for the sake of illustration that soft final-state muons with energies $E_{\mu,\textrm{out}}> 10~\textrm{GeV}$ can be identified. We also require $E_{\mu,\textrm{out}} < 50~\textrm{GeV}$ to suppress SM backgrounds.
\medskip

\textbf{Tau pair invariant mass} The muon energy loss cut does not strictly require a precise measurement of final-state energy, provided the value remains below the $50~\textrm{GeV}$ threshold. However, such a measurement would further constrain the tau pair invariant mass, $m_{\tau\tau}$, given the negligible nuclear recoil in the coherent scatterings. This reconstruction would be further enhanced by the measurement of charged tau decay products and intermediate tau tracks.

\begin{figure*}
 \centering
    \includegraphics[width=0.49\linewidth]{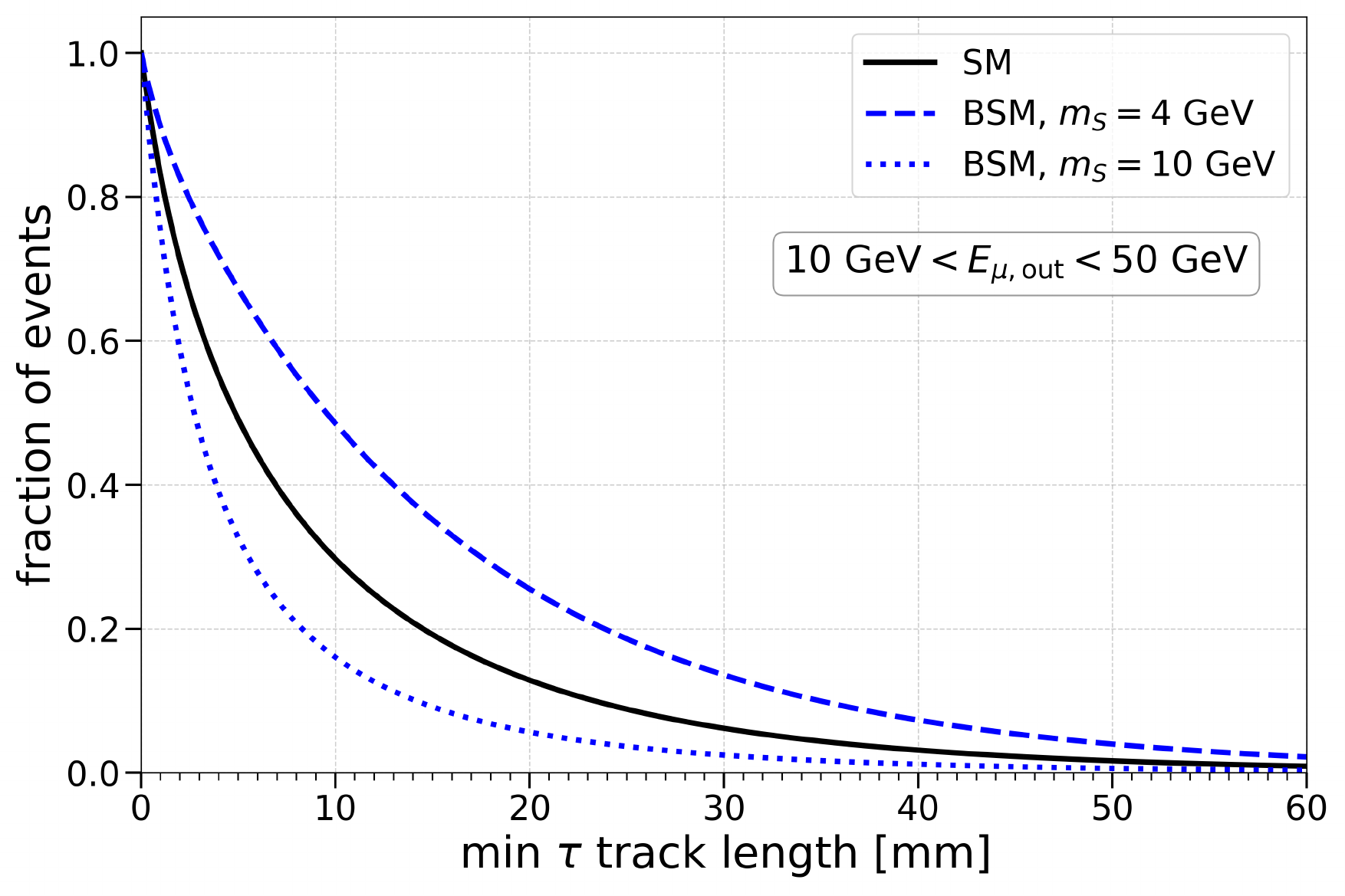}
    \includegraphics[width=0.49\linewidth]{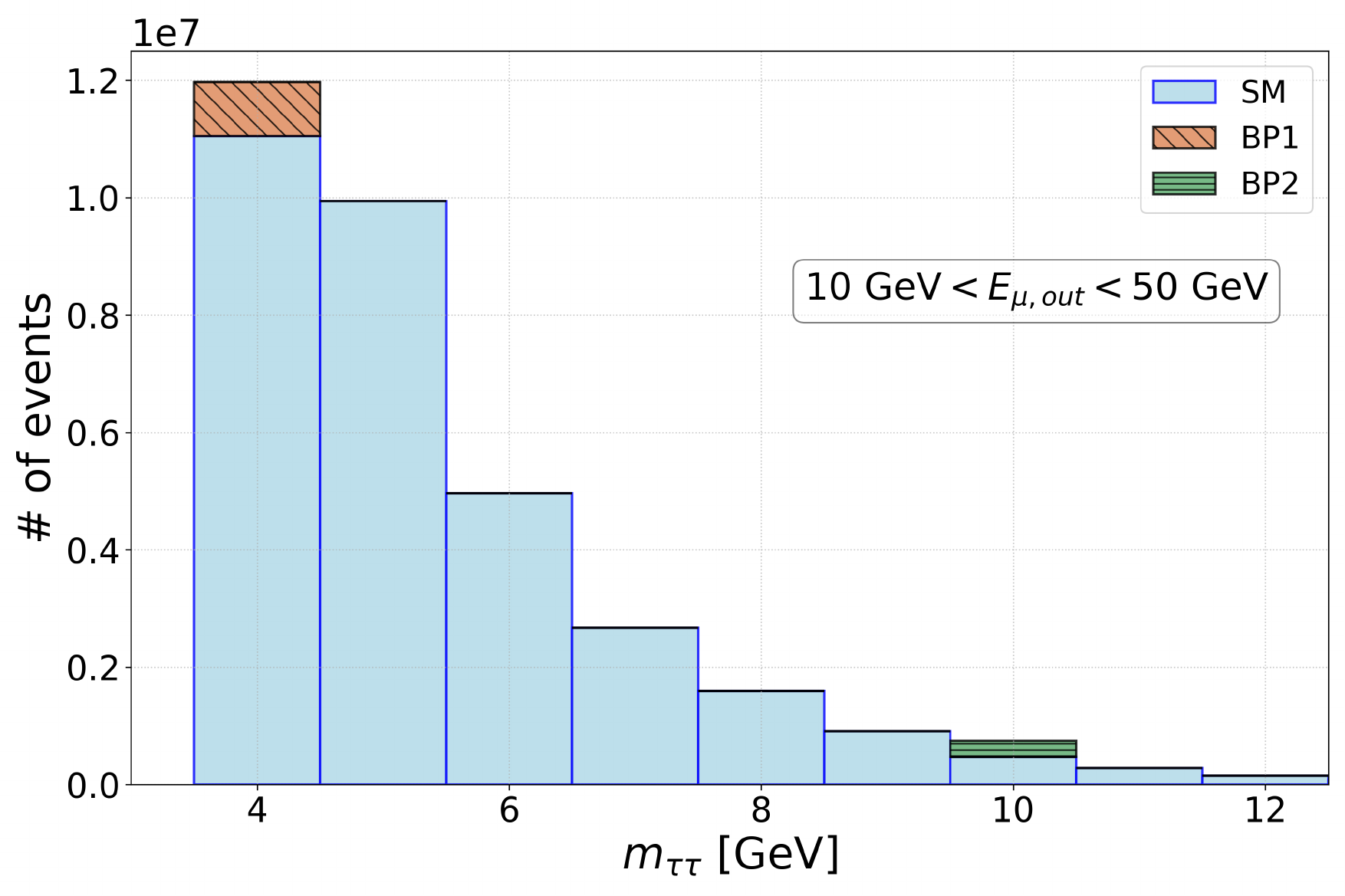}
    \caption{Kinematic distributions for muon-induced di-tau production in HEMERA operating at $E_\mu = 1~\textrm{TeV}$ with $10^{18}$ MOT. \textsl{Left:} The fraction of events where both final-state tau leptons have a track length exceeding the threshold indicated on the horizontal axis. \textsl{Right:} The invariant mass distribution of the tau pair, $m_{\tau\tau}$, comparing the SM background with the BSM signal benchmarks BP1 ($m_S=4~\textrm{GeV}$) and BP2 ($m_S=10~\textrm{GeV}$). Both panels assume the muon energy loss cut $10~\textrm{GeV} < E_\mu,{\textrm{out}} < 50~\textrm{GeV}$ has been applied. The BSM signals are shown as stacked histograms on top of the SM background, assuming an invariant mass resolution of $\Delta m_{\tau\tau}=1~\textrm{GeV}$.}
    \label{fig:invmass}
\end{figure*}

In the left panel of \cref{fig:invmass}, we show the fraction of di-tau production events where both tau track lengths exceed the minimal threshold indicated on the horizontal axis. Results are presented for both SM events and BSM scalar production followed by the prompt $S\to\tau^+\tau^-$ decay. The plot assumes a $1~\textrm{TeV}$ incident muon energy and includes events satisfying the muon energy loss cut, $10~\textrm{GeV}\lesssim E_{\mu,\textrm{out}}\lesssim 50~\textrm{GeV}$, where the lower limit represents the detectability criterion for soft final-state muons discussed above. As can be seen, typical boosted tau decay lengths in this case are of the order of a few millimeters; therefore, facilitating successful intermediate track identification in modern electronic detectors with sub-millimeter position resolution to be utilized in HEMERA's high-granularity silicon planes, cf. \cref{sec:experiment}. 

We note that applying the muon energy loss cut effectively restricts the energy of the outgoing scalar to $E_S\approx (950-990)~\textrm{GeV}$. The resulting tau lepton energy distribution from $S$ decays exhibits a characteristic box shape in the lab frame, with a width of $\Delta E_\tau\simeq E_S\,\beta_\tau^\ast$. Here, we assume $E_S\simeq p_S$ for a highly boosted scalar, and $\beta_\tau^\ast = \sqrt{1-4\,m_\tau^2/m_S^2}$ is the tau lepton velocity in the $S$ rest frame. As $\beta_\tau^\ast$ increases with $m_S$, so does the width of the tau energy distribution. The production of softer tau leptons in this regime leads to a characteristic reduction in the minimum $\tau$ track length for $m_S=10~\textrm{GeV}$ (BP2) compared to $m_S=4~\textrm{GeV}$ (BP1), as shown in the plot.

Due to the significant boost of the BSM scalar $S$, the two tau tracks produced in its decay will exhibit a very small opening angle, $\theta_{\tau\tau}\sim m_S/E_S\sim 10~\textrm{mrad}$ for $m_S = \mathcal{O}(10~\textrm{GeV})$ and $E_S\sim 1~\textrm{TeV}$. Consequently, the two intermediate tau tracks will likely appear as a single ``fat'' (or double) track within the detector. Specifically, a tau decay length of approximately $10~\textrm{mm}$ is required to achieve a spatial separation of $10~\textrm{mrad}\times 10~\textrm{mm} \simeq 100~\mu m$, which is comparable to the pitch size of a typical silicon tracker. For shorter track lengths where spatial separation is not yet possible, these collimated tau leptons can be discriminated from single-particle tracks by measuring the ionization energy loss ($dE/dx$) in the tracker.

The precision of the invariant mass measurement depends on the tau decay topology. Specifically, whether the taus undergo 1-prong (single charged track) or 3-prong (three charged tracks) decays. For the dominant 1-prong decays, which have a branching fraction of $B_1\simeq 85\%$~\cite{ParticleDataGroup:2024cfk}, identifying the exact decay vertex positions is challenging because only one visible charged track is produced per tau lepton. In this scenario, the total energy of the di-tau pair can be estimated via the muon energy loss. For illustration, assuming an incident muon energy spread of $\mathcal{O}(0.1\%)$ and a few-percent energy resolution for the soft outgoing muon~\cite{InternationalMuonCollider:2025sys}, the di-tau energy could be determined with an accuracy of several GeV. This represents sub-percent precision given that $E_{\tau^+}+E_{\tau^-} \gtrsim 950~\textrm{GeV}$ after applying the muon energy loss cut.

The energy splitting between the two tau leptons can then be determined by analyzing the charged tracks from the $\tau$ decays. In the high-boost regime, assuming the charged tracks are effectively collinear with their parent tau leptons, the energy ratio can be approximated as $E_{\tau^+}/E_{\tau^-} \simeq \theta_{\tau^-}/\theta_{\tau^+}$, where $\theta_{\tau^-}$ and $\theta_{\tau^+}$ are the angles relative to the parent $S$ momentum. In the HEMERA high-precision silicon active target environment, an angular resolution of $\sigma_\theta \sim 100~\mu\textrm{rad}$ is achievable~\cite{ParticleDataGroup:2024cfk}, yielding a $2\%$ accuracy for typical angles of $\theta_i\sim 5~\textrm{mrad}$. Propagating these uncertainties results in a $3\%$ uncertainty on the energy ratio $E_{\tau^+}/E_{\tau^-}$.

In this configuration, the measurement of individual tau energies is primarily limited by the energy resolution of the muon spectrometer and the angular resolution of the tracker. By combining the sub-percent precision of the di-tau energy with the $\mathcal{O}(1\%)$ precision on the opening angle $\theta_{\tau^+\tau^-}$ and the constrained energy ratio $E_{\tau^+}/E_{\tau^-}$, an invariant mass resolution $\Delta m_{\tau\tau}\lesssim 1~\textrm{GeV}$ could be achieved for $m_S \sim 10~\textrm{GeV}$, where $m_{\tau\tau}\simeq \sqrt{2m_\tau^2+E_{\tau^+}E_{\tau^-}\,\theta_{\tau^+\tau^-}^2}$ at high energies. 

For the 3-prong $\tau$ lepton decay mode, which accounts for approximately $B_3\simeq 15\%$ of the tau branching fraction~\cite{ParticleDataGroup:2024cfk}, the reconstruction of $m_{\tau\tau}$ could be further improved. The primary challenge in this case arises from the aforementioned ``fat track'' effect; at $>100~\textrm{GeV}$ energies, the three collimated daughter charged tracks create a high-occupancy environment within a few milliradians. As mentioned above, however, this high track multiplicity may also provide an additional experimental signature via ionization energy loss ($dE/dx$). In the extreme case of a 3+3 prong topology where the two taus have not yet spatially diverged, the detector would observe a single ``fat'' track with a $dE/dx$ signature of up to six minimum ionizing particles. This would provide a strong auxiliary handle for identifying the di-tau system and rejecting single-particle backgrounds even before individual tracks are resolved.

The sub-millimeter spatial resolution of the silicon tracker will eventually allow these collimated bundles to be resolved into individual tracks. In 1+3 or 3+3 prong events, the intersection of these tracks enables a more precise localization of the secondary tau decay vertex. By connecting the production vertex to this secondary vertex, the direction of the parent tau lepton can be determined with a higher geometric certainty than in 1-prong decays, where the single track is used as a proxy for the tau direction. This reduces the uncertainty in the opening angle $\theta_{\tau\tau}$ and the energy ratio $E_{\tau^+}/E_{\tau^-}$, further mitigating the energy information lost to neutrinos. Considering events with at least one 3-prong decay could then push the $\Delta m_{\tau\tau}$ resolution below the $1~\textrm{GeV}$ threshold and provide a clearer experimental signature of the di-tau production.

For simplicity, we assume a fixed resolution of $\Delta m_{\tau\tau} = 1~\textrm{GeV}$ in the HEMERA analysis below. In the right panel of \cref{fig:invmass}, we show the $m_{\tau\tau}$ distributions for the SM di-tau production events and the stacked signal statistics for BSM signal benchmarks BP1 and BP2. The latter are primarily localized in individual $m_{\tau\tau} \simeq m_S$ bins, while SM backgrounds peak close to the di-tau mass threshold with only a tail of the distribution reaching to larger $m_{\tau\tau}$. The distributions have been obtained after applying the muon energy loss cut discussed above. As can be seen, considering the cut on the tau pair invariant mass allows for significantly improving detection prospects for new scalars. This is especially true for increasing $S$ mass, for which the BSM signal could easily be swamped by the dominant SM background had the $m_{\tau\tau}$ bound not been considered.

\begin{figure}
 \centering
    \includegraphics[width=\linewidth]{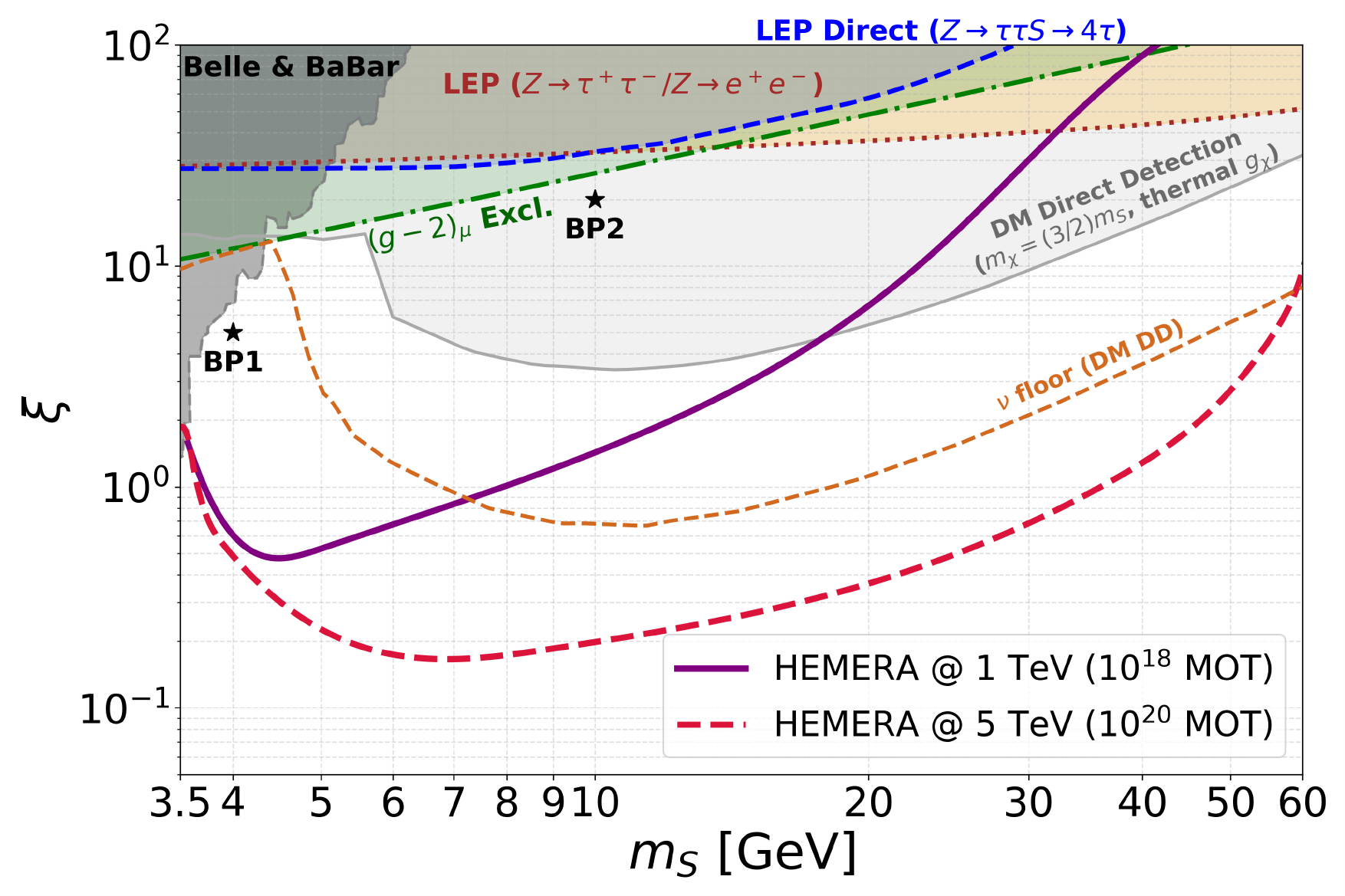}
    \caption{The projected 95\% CL exclusion limit on $\xi$ versus $m_S$ for the proposed HEMERA setup with $10^{18}$ MOT for 1~TeV incident muons (purple solid line). The red dashed line represents the sensitivity for $10^{20}$ MOT with 5~TeV muons. We also mark the benchmark points BP1 and BP2 discussed in the text. The light gray shaded region represents the loop-induced exclusion bounds from dark matter direct detection searches by the LZ~\cite{LZ:2024zvo} and XENONnT~\cite{XENON:2026qow} experiments. These bounds, alongside the corresponding neutrino floor~\cite{OHare:2021utq} (dashed brown line), are derived assuming a dark matter mass $m_\chi = (3/2)\,m_S$ and a dark coupling $g_\chi$ fixed to reproduce the DM thermal relic density. Also shown are constraints from BaBar~\cite{BaBar:2020jma}, Belle~\cite{Belle:2022gbl} and Belle~II~\cite{Belle-II:2023ydz}, the bound from $(g-2)_\mu$ from eq.~(\ref{eq:amubsm}), the indirect constraints from LEP precision measurements of lepton flavor universality in $Z \to \ell\ell$ decays~\cite{ALEPH:2005ab,Chun:2016hzs}, and the bound from the direct searches for $e^+e^-\to \tau\tau S \to 4\tau$~\cite{DELPHI:2004bco}.}
    \label{fig:constraints}
\end{figure}

\textbf{Sensitivity} In \cref{fig:constraints}, we present the projected $95\%$ CL exclusion limits for the proposed HEMERA experiment. Assuming $10^{18}$ MOT at $1~\textrm{TeV}$ on a $10~\textrm{kg}$ target, HEMERA significantly constrains the parameter space of the considered BSM model. Furthermore, a full MAP physics run with $10^{20}$ MOT and a $5~\textrm{TeV}$ muon beam, shown by the red dashed line, would push the sensitivity to even smaller couplings, potentially reaching $\xi \sim 0.1$ in this mass range. In the plot, we also explicitly mark the aforementioned benchmark points BP1 ($m_S = 4~\textrm{GeV}$, $\xi = 5$) and BP2 ($m_S = 10~\textrm{GeV}$, $\xi = 20$). These projections are compared with existing bounds from searches for leptophilic scalars at BaBar~\cite{BaBar:2020jma}, Belle~\cite{Belle:2022gbl}, and Belle II~\cite{Belle-II:2023ydz}; see also Ref.~\cite{Alda:2024cxn} for a further discussion.

We note that the BSM sensitivity will ultimately be constrained by the systematic uncertainties on the predicted number of di-tau production events in the detector. A resonance search focused on the expected bump in the tau invariant mass would suppress uncertainties in SM event rates related to, e.g., the total muon flux or the SM cross section prediction. This can be done by measuring the SM rates in neighboring bins in $m_{\tau\tau}$, where no BSM events are expected for a given leptophilic scalar mass. While this data-driven background rejection will significantly improve the prospects for the BSM search, other systematic uncertainties, such as those related to invariant mass resolution and shape modeling, will limit the true experimental reach. We leave detailed analyses of such effects for future studies. 

In \cref{fig:constraints}, we also show the updated $95\%$ CL exclusion bound on BSM contributions to the muon's anomalous magnetic moment. Incorporating the latest experimental measurements~\cite{Muong-2:2006rrc,Muong-2:2021ojo,Muong-2:2023cdq,Muong-2:2025xyk} alongside the 2025 SM theoretical update~\cite{Aliberti:2025beg}, we adopt the discrepancy $a_\mu^{\text{exp}} - a_\mu^{\text{SM}} = 38(63)\times 10^{-11}$. This leads to a one-sided $95\%$ CL upper limit on any BSM contribution of $\Delta a_\mu^{\textrm{BSM}}< 1.42\times 10^{-9}$. For the scalar particle $S$ considered here, the one-loop contribution to $(g-2)_\mu$ is calculated as~\cite{Batell:2016ove}
\begin{equation}
\label{eq:amubsm}
\Delta a_\mu^{\textrm{BSM}} = \frac{\xi^2}{8\pi^2} \frac{m_\mu^2}{v^2} \int_0^1 \frac{x^2(2-x)}{x^2 + (1-x)(m_S/m_\mu)^2} dx\ .
\end{equation}
Coupling strengths $\xi$ that yield a contribution exceeding the $1.42\times 10^{-9}$ threshold are excluded, as represented by the shaded region in the parameter space. Furthermore, we assess the theoretical consistency of the model by requiring the tau Yukawa coupling,  $y_\tau=\xi m_\tau/v$, to remain within the perturbative regime. Taking the simple criterion $y_\tau<\sqrt{4\pi}$, we find a mass-independent perturbativity bound of $\xi\lesssim 492$. As shown in the results, this theoretical ceiling is significantly less restrictive than the constraints imposed by $(g-2)_\mu$ over the entire $m_S$ range of interest.

Notably, the BSM scalar of our interest can similarly contribute to the anomalous magnetic moment of the tau lepton, $a_\tau$. While current bounds on $a_\tau$ are not yet competitive with the muon anomaly~\cite{DELPHI:2003nah,ATLAS:2022ryk,CMS:2022arf}, future Belle-II measurements are projected to constrain new physics contributions to this quantity at the level of $|a^{\textrm{NP}}_\tau| < 1.75\times 10^{-5}$~\cite{Chen:2018cxt}. A proposed polarization upgrade of the SuperKEKB collider could result in even stronger sensitivity, potentially reaching the $10^{-6}$ level~\cite{Crivellin:2021spu}. Given the Yukawa-like coupling of the new scalar $S$, the BSM contribution is significantly enhanced for the $\tau$ lepton relative to the muon. This enhancement arises from the $m_\ell^2$ scaling in the prefactor of \cref{eq:amubsm} and a corresponding reduction in the integrand suppression when substituting $(m_S/m_\mu)^2\to (m_S/m_\tau)^2$. Based on the static limit, one would expect future sensitivity to reach $\xi\sim 10-15$ for $m_S=10~\textrm{GeV}$. We note, however, that Belle-II will measure this contribution dynamically at the $\Upsilon(4S)$ resonance, $\sqrt{s}\approx 10.58~\textrm{GeV}$. In this regime, the measured $\tau^+\tau^-\gamma^\ast$ vertex form factor $F_2(s)$ exhibits a power-law dependence on the energy. This would effectively weaken the sensitivity to $\xi$ compared to the static prediction, an estimate of which we leave for a future study. 

Complementary to bounds from $\delta a_\mu$, the leptophilic scalar $S$ induces 1-loop virtual corrections to the $Z\tau\tau$ vertex, which are strongly constrained by precision electroweak measurements on the $Z$ boson resonance at LEP. The most stringent bound arises from the lepton flavor universality test in the ratio of the $Z$ leptonic partial widths, $\Gamma(Z \to \tau\tau)/\Gamma(Z \to ee) = 1.0019 \pm 0.0032$~\cite{ALEPH:2005ab}.

The relevant BSM-driven shift in the $Z \to \tau\tau$ partial width is governed by the interference between the SM tree-level amplitude and the 1-loop scalar exchange, given by
\begin{align}
    \label{eq:delta_tautau}
    \delta_{\tau\tau} & \equiv \frac{\Gamma(Z \to \tau\tau)}{\Gamma(Z \to ee)} - \left[ \frac{\Gamma(Z \to \tau\tau)}{\Gamma(Z \to ee)} \right]_{\text{SM}} \\
    & \simeq \left[ \frac{\Gamma(Z \to \tau\tau)}{\Gamma(Z \to ee)} \right]_{\text{SM}} \times\frac{2 g_L^e \text{Re}(\delta g_L^S) + 2 g_R^e \text{Re}(\delta g_R^S)}{(g_L^e)^2 + (g_R^e)^2},\nonumber
\end{align}
where $g_L^e = -1/2 + \sin^2\theta_W$ and $g_R^e = \sin^2\theta_W$ are the SM electron chiral couplings, and we neglect the strongly suppressed BSM contributions to the $Z\to ee$ decay width, $\Gamma(Z \to ee)\simeq \Gamma_{\textrm{SM}}(Z \to ee)$. Considering the mass of the $\tau$ lepton, the SM prediction for the ratio of the decay widths is $[\Gamma(Z \to \tau\tau)/\Gamma(Z \to ee)]_{\textrm{SM}}\simeq 0.9977$~\cite{ALEPH:2005ab}. The effective couplings for the Yukawa-like leptophilic scalar, retaining only the finite parts of the loop integrals to treat the introduction of the BSM scalar field as an effective theory, yield
\begin{eqnarray}
    \delta g_L^S &=& \Xi \left\{ -\frac{1}{2} B_Z(r_S) + \sin^2\theta_W \left[ B_Z(r_S) + \tilde{C}_Z(r_S) \right] \right\}, \nonumber \\
    \delta g_R^S &=& \Xi \left\{ -\frac{1}{2} \tilde{C}_Z(r_S) + \sin^2\theta_W \left[ B_Z(r_S) + \tilde{C}_Z(r_S) \right] \right\}, \nonumber\\\label{eq:dgR}
\end{eqnarray}
where $r_S = m_S^2/m_Z^2$, $\Xi=\frac{\xi^2 m_\tau^2}{16\pi^2 v^2}$, while $B_Z$ and $\tilde{C}_Z$ represent the finite combinations of Passarino-Veltman two-point and three-point scalar loop integrals. In specific UV completions, additional finite terms are expected to appear once the heavy fields are integrated out, cf. the discussion of the Two Higgs Doublet Model (2HDM)~\cite{Chun:2016hzs}. Using the measured LEP central value and its uncertainty, alongside the aforementioned SM expectation of $[\Gamma(Z \to \tau\tau)/\Gamma(Z \to ee)]_{\textrm{SM}}$, we require the BSM contribution to lie within the 95\% CL interval $-0.0021 \lesssim \delta_{\tau\tau} \lesssim 0.0105$, and we plot the resulting constraint in \cref{fig:constraints}.

Beyond virtual loop effects, direct bounds arise from LEP searches for multi-tau final states, specifically $e^+e^-\to \tau^+\tau^-S\to 4\tau$~\cite{DELPHI:2004bco}. We map the corresponding coupling reduction factor to our parameter as $C_{\tau\tau(h\to\tau\tau)}\simeq \xi$, assuming a nearly $100\%$ branching fraction of the BSM scalar to tau pairs, and include this constraint in \cref{fig:constraints}.

While recent LHC analyses also target $4\tau$ signatures~\cite{CMS:2026ntm}, the relevant event selection assumes the intermediate production of two heavy states with masses of order tens of GeV or so. These searches rely on the resulting large total transverse mass ($m_T^{tot}$) and high-$p_T$ tau leptons to discriminate the signal from SM backgrounds. In contrast, in our case, the signal would be dominated by the $\tau$-strahlung of a BSM scalar, $pp\to Z^\ast/\gamma^\ast\to\tau\tau S\to 4\tau$, which produces a softer, low-$m_T^{tot}$ signature. As a result, these bounds are not directly applicable to our scenario and would require a dedicated analysis.

Finally, it is instructive to consider the complementarity of the proposed HEMERA search with direct DM detection. In \cref{fig:constraints}, we show the leading constraints from recent DM direct detection experiments, specifically LZ~\cite{LZ:2024zvo} and XENONnT~\cite{XENON:2026qow}. To map these bounds onto the $(m_S, \xi)$ parameter space, we calculate the loop-induced DM-nucleon scattering cross section via two-photon exchange~\cite{Garani:2021ysl}. For concreteness, we fix the DM mass to $m_\chi = (3/2)\,m_S$, which ensures the visible decay $S \to \tau^+\tau^-$, and set the dark coupling $g_\chi$ to obtain the thermal relic density of $\Omega_\chi h^2 \simeq 0.12$~\cite{Planck:2018vyg} via secluded annihilations, $\chi\bar{\chi}\to SS$. Under these specific assumptions, the loop-induced direct detection cross section places a stringent bound, shown by the light gray shaded region in \cref{fig:constraints}, which exceeds other bounds for $m_S$ in between a few GeV and several tens of GeV. 

In addition to current experimental limits, we also display the neutrino floor (dashed line) in \cref{fig:constraints} following Ref.~\cite{OHare:2021utq}, which we map to the $(m_S, \xi)$ plane under the same benchmark assumptions. This represents the boundary where direct detection experiments will be overwhelmed by the irreducible background from coherent neutrino-nucleus scattering. As shown in the plot, for lighter DM masses, represented by the BP1 benchmark, the projected sensitivity of HEMERA extends beyond this limit, where traditional direct detection searches face fundamental limitations. This clearly illustrates the complementarity of the proposed accelerator search, which can probe the thermal DM parameter space where underground detectors lose their primary sensitivity.

Instead, in this specific scenario, the benchmark point BP2 is already presently excluded. It is crucial to emphasize, however, that these direct detection limits are sensitive to additional assumptions about the dark sector. If the DM relic abundance is generated non-thermally (e.g., via the freeze-in mechanism), or if the mass hierarchy deviates from this specific choice, the required dark coupling can be smaller, and the constraints can be relaxed. In contrast, the HEMERA projections depend exclusively on the scalar's visible couplings to SM leptons, providing a robust discovery channel for the mediator itself.

As demonstrated in \cref{fig:constraints}, the proposed HEMERA setup would yield highly competitive constraints, with a reach extending to $\xi\sim 0.1$ for $m_S=10~\textrm{GeV}$. In this mass range, these projections are significantly more stringent than anticipated sensitivities from the ILC ($\xi\gtrsim \mathcal{O}(1)$)~\cite{Chun:2019sjo,Chun:2021rtk,Jia:2021mwk} and are comparable to the reach of a future Tera-Z factory at FCC-ee~\cite{Dam:2018rfz}. HEMERA thus provides an independent test of the leptophilic BSM scenario, with the potential to eventually outperform future circular $e^+e^-$ machines by scaling up the detector volume.

\section{Conclusion}
\label{sec:conclusion}

In this work, we have demonstrated that muon-induced di-tau production, $\mu A\to \mu\tau^+\tau^-A$, serves as a highly sensitive and clean probe of new physics at future high-energy muon facilities. By utilizing the characteristic energy loss of the primary muon in conjunction with the identification of short, boosted $\tau$ tracks and the measurement of the tau pair invariant mass, we have shown that the Standard Model background can be effectively controlled. For a leptophilic scalar mediator $S$ with Yukawa-like couplings, the projected sensitivity of a TeV-scale experiment, even with a relatively compact target, surpasses existing constraints from $B$-factories and the muon anomalous magnetic moment by more than an order of magnitude in the few-GeV mass range. This highlights the trident process as a primary discovery channel for new mediators that prefer third-generation lepton couplings.

A particularly compelling motivation for this search is its ability to probe the dark sector through the thermal WIMP-like DM portal. In scenarios where the leptophilic scalar mediates interactions between the SM and a DM candidate $\chi$, the resulting thermal relic density is typically determined by the secluded annihilation $\bar{\chi}\chi\to SS$. While such models lack tree-level couplings to quarks and feature suppressed electron scattering rates, exact calculations of loop-induced processes show that specific thermal setups can still be strongly constrained by current direct detection experiments. However, these bounds remain highly sensitive to assumptions regarding the dark sector mass hierarchy and the cosmological origin of the relic abundance. We have shown that the proposed di-tau trident measurement provides a promising way to bridge this gap by directly targeting the mediator. Crucially, in the mass range $m_S<m_\chi$  in scenarios with thermal relic $\chi$, HEMERA can  probe the leptophilic scalar coupling and scalar mass in regions of parameter space beyond the neutrino floor, which is the fundamental limit where underground detectors are overwhelmed by irreducible backgrounds to direct detection. With indirect detection probes also remaining challenging for $p$-wave suppressed secluded annihilation modes, this accelerator-based search offers an independent tool to explore this scenario.

While we focused on a leptophilic scalar as a representative BSM model, the proposed search strategy could apply to a wider range of new physics scenarios. The muon-induced di-tau final state is sensitive to light, weakly coupled new physics that preferentially interacts with second- and third-generation leptons. A notable example is a $Z'$ gauge boson arising from the anomaly-free $U(1)_{L_\mu - L_\tau}$ model~\cite{Altmannshofer:2014cfa}. On top of this, the muon trident signature can indirectly probe much heavier new physics, e.g., leptoquarks or heavy vector-like leptons that can generate dimension-$6$ electroweak dipole operators that modify the $\gamma\tau\tau$ vertex~\cite{Aebischer:2021uvt}, thereby indirectly varying the predicted di-tau kinematic distributions.

The experimental feasibility of this search is underscored by the remarkable intensity of future muon beams. We have introduced the concept of the High-Energy Muon Electronic Research Apparatus (HEMERA), an active-target detector capable of performing a world-leading search by leveraging the enormous flux of high-energy muons. Importantly, HEMERA’s active-target operation mode is highly complementary to previously proposed muon beam-dump searches. While beam dumps are optimized for long-lived particles, HEMERA’s high-granularity target would enable the reconstruction of prompt or short-lived BSM signatures that would otherwise be lost.

An integrated exposure of $10^{18}$ muons on target already provides the high statistics necessary to probe rare BSM processes with high precision using even a compact detector setup. Reaching an ultimate sensitivity, either through a larger detector mass or a longer physics run yielding up to $10^{20}$ MOT, would push the discovery reach to even smaller couplings. This experimental program is ideally suited for the preparatory phase of the future Muon Collider involving individual muon beams, and its utility extends to the full collider facility. Ultimately, the active-target detector concept and the high-energy muon trident process offer a versatile and powerful tool for exploring the dark sector, further motivating the development of TeV-scale muon beam infrastructure. Beyond the signatures discussed here, the potential of the active-target HEMERA detector operating in high-energy muon beams should be further explored for a wider range of new physics applications and precision SM measurements.\\\\
\section*{Acknowledgements}
We thank Brian Batell and Felix Kling for useful discussions. We thank the authors of Ref.~\cite{Graziani:2026noc} for useful comments on dark matter direct detection constraints on the BSM model presented here. The University of Iowa group is supported in part by US DOE grant DE-SC-0010113. ST is supported by the National Science Centre, Poland (research grant No. 2021/42/E/ST2/00031). ST is also partially supported by Teaming for Excellence grant Astrocent Plus (GA: 101137080) funded by the European Union, with complementary national funding from the MNiSW (MNiSW/2025/DIR/811). 

\bibliographystyle{apsrev4-1}
\bibliography{lib}

\end{document}